\documentclass[12pt]{article}
\def\be{\begin{equation}}
\def\ee{\end{equation}}
\def\bea{\begin{eqnarray}}
\def\eea{\end{eqnarray}}
\usepackage{graphicx}

\catcode`\@=11
\def\lsim{\mathrel{\mathpalette\@versim<}}
\def\gsim{\mathrel{\mathpalette\@versim>}}
\def\@versim#1#2{\vcenter{\offinterlineskip
\ialign{$\m@th#1\hfil##\hfil$\crcr#2\crcr\sim\crcr } }}
\catcode`\@=12
\input epsf.tex

\catcode`@=12
\begin{document}
\thispagestyle{empty}
\begin{flushright}
UCRHEP-T505\\
June 2011\
\end{flushright}
\vspace{0.3in}
\begin{center}
{\LARGE \bf Nonzero $\theta_{13}$ for neutrino mixing\\
 in the context of $A_4$ symmetry\\}
\vspace{1.2in}
{\bf Ernest Ma and Daniel Wegman\\}
\vspace{0.2in}
{\sl Department of Physics and Astronomy, University of California,\\ 
Riverside, California 92521, USA\\}
\end{center}
\vspace{1.0in}
\begin{abstract}\
In the original 2004 paper which first derived tribimaximal mixing in the 
context of $A_4$, i.e. the non-Abelian finite symmetry group of the 
tetrahedron, as its simplest application, it was also pointed out how 
$\theta_{13} \neq 0$ may be accommodated.  On the strength of the new T2K 
result that $0.03 (0.04) \leq \sin^2 2 \theta_{13} \leq 0.28 (0.34)$ for 
$\delta_{CP}=0$ and normal (inverted) neutrino mass hierarchy, we perform 
a more detailed analysis of how this original idea may be realized in 
the context of $A_4$.
\end{abstract}

\newpage
\baselineskip 24pt

Neutrino oscillations require nonzero neutrino masses as well as nonzero 
neutrino mixing angles.  The current combined world data 
imply~\cite{pdg10}
\begin{eqnarray}
&& 7.05 \times 10^{-5} ~{\rm eV}^2 \leq \Delta m^2_{21} \leq 8.34 \times 10^{-5} 
~{\rm eV}^2, \\
&& 2.07 \times 10^{-3} ~{\rm eV}^2 \leq \Delta m^2_{32} \leq 2.75 \times 10^{-3} 
~{\rm eV}^2, \\
&& 0.36 \leq \sin^2 \theta_{23} \leq 0.67, ~~~ 
0.25 \leq \sin^2 \theta_{12} \leq 0.37,\\ 
&& \sin^2 \theta_{13} \leq 0.035 ~(90\%~{\rm CL}). 
\end{eqnarray}
However, the T2K Collaboration recently announced that a new  
measurement~\cite{t2k11} has yielded a nonzero $\theta_{13}$ at 90\% 
confidence level, i.e.
\begin{equation}
0.03 (0.04) \leq \sin^2 2 \theta_{13} \leq 0.28 (0.34)
\end{equation}
for $\delta_{CP}=0$ and normal (inverted) neutrino mass hierarchy. 

For several years now, the mixing matrix $U_{l\nu}$ linking the charged 
leptons $(e, \mu, \tau)$ to the neutrino mass eigenstates 
$(\nu_1, \nu_2, \nu_3)$ has often been assumed to be of tribimaximal 
form~\cite{hps02}, i.e.
\begin{equation}
U_{TB} = \pmatrix{\sqrt{2/3} & 1/\sqrt{3} & 0 \cr -1/\sqrt{6} & 1/\sqrt{3} 
& -1/\sqrt{2} \cr -1/\sqrt{6} & 1/\sqrt{3} & 1/\sqrt{2}},
\end{equation}
which predicts $\theta_{13}=0$.  This is particularly appealing because 
it was derived in 2004~\cite{m04} from the simple application of the symmetry 
group $A_4$, first used for understanding maximal $\nu_\mu-\nu_\tau$ 
mixing in 2001~\cite{mr01}. However, even in that original 2004 
paper~\cite{m04}, the possibility of $\theta_{13} \neq 0$ was already 
anticipated.  Although the new T2K result~\cite{t2k11} is only 2.5$\sigma$ 
away from zero, it is the most solid experimental indication to date of 
this possibility. Here we offer a more detailed analysis of how $\theta_{13} 
\neq 0$ may be realized in the context of $A_4$.

As is well-known, $A_4$ is the group of the even permutation of 4 objects.  
It is also the symmetry of the perfect three-dimensional 
tetrahedron~\cite{m02}.  It has 12 elements 
and 4 irreducible representations: \underline{1}, \underline{1}$'$, 
\underline{1}$''$, \underline{3}, with the multiplication rule 
\begin{equation}
\underline{3} \times \underline{3} = \underline{1} + \underline{1}' + 
\underline{1}'' + \underline{3} + \underline{3}. 
\end{equation}
The first step in understanding neutrino mixing is to show that $A_4$ allows 
the charged-lepton mass matrix 
to be diagonalized by the Cabibbo-Wolfenstein matrix~\cite{c78,w78}
\begin{equation}
U_{CW} = {1 \over \sqrt{3}} \pmatrix{1 & 1 & 1 \cr 1 & \omega & \omega^2 \cr 
1 & \omega^2 & \omega}
\end{equation}
where $\omega = e^{2 \pi i/3} = -1/2 + i \sqrt{3}/2$, with three independent 
eigenvalues, i.e. $m_e, m_\mu, m_\tau$.  This has been achieved in two ways. 
One is the original proposal of 2001~\cite{mr01}.  The other was discovered 
later in 2006~\cite{m06}.  In the former, the lepton assignments are 
$L_i = (\nu_i,l_i) \sim \underline{3}, ~l^c_1 \sim \underline{1}, 
~l^c_2 \sim \underline{1}', ~l^c_3 \sim \underline{1}''$, with 3 
Higgs doublets $\Phi_i = (\phi^0_i, \phi^-_i) \sim \underline{3}$. 
In the latter, they are $L_i = (\nu_i,l_i) \sim \underline{3}, 
~l^c_i \sim \underline{3}$, with 4 Higgs doublets 
$\Phi_i = (\phi^0_i, \phi^-_i) \sim \underline{3}, ~\Phi_0 \sim \underline{1}$. 
Assuming $v_1 = v_2 = v_3$ for the vacuum expectation values of $\Phi_i$, 
which correspond to a $Z_3$ residual symmetry 
(lepton triality)~\cite{m09,m10-1,ckmo11,cdmw11}, the seemingly impossible 
result of a diagonal charged-lepton matrix is always obtained from $U_{CW}$ 
of Eq.(8), independent of the values of $m_e, m_\mu, m_\tau$.  This is 
a highly nontrivial result, which motivates how the otherwise arbitrary 
$3 \times 3$ neutrino mass should be organized.  It argues strongly for 
an underlying non-Abelian symmetry with a three-dimensional irreducible 
representation, the smallest of which is $A_4$.

We now consider the neutrino mass matrix in the original $A_4$ basis. 
Let there be 6 heavy Higgs triplets~\cite{ms98}:
\begin{equation}
\xi_1 \sim \underline{1}, ~~ \xi_2 \sim \underline{1}', ~~ 
\xi_3 \sim \underline{1}'', ~~ 
\xi_i \sim \underline{3} ~~(i=4,5,6),
\end{equation}
where $\xi_i = (\xi^{++}_i,\xi^+_i,\xi^0_i)$. Then
\begin{equation}
{\cal M}_\nu = \pmatrix{a + b + c & f & e \cr f & a + \omega b + \omega^2 c & 
d \cr e & d & a + \omega^2 b + \omega c},
\end{equation}
where $a$ comes from $\langle \xi^0_1 \rangle$, 
$b$ from $\langle \xi^0_2 \rangle$, 
$c$ from $\langle \xi^0_3 \rangle$, 
$d$ from $\langle \xi^0_4 \rangle$, 
$e$ from $\langle \xi^0_5 \rangle$, 
$f$ from $\langle \xi^0_6 \rangle$.
As it stands, there is of course no prediction at all.  For a pattern to 
emerge, the way $A_4$ breaks into its subgroups must be considered. 
For $b=c$ and $e=f=0$, which breaks $A_4$ to $Z_2$, the neutrino mass 
matrix, written in the basis where the charged-lepton mass matrix is 
diagonal, is given by
\begin{equation}
{\cal M}_\nu^{(e,\mu,\tau)} = U_{CW}^\dagger {\cal M}_\nu U_{CW}^* 
= \pmatrix{a+(2d/3) & b-(d/3) & b-(d/3) \cr b-(d/3) & b+(2d/3) & a-(d/3) \cr 
b-(d/3) & a-(d/3) & b+(2d/3)},
\end{equation}
which is indeed diagonalized by $U_{TB}$ of Eq.(6), with eigenvalues 
$m_1 = a-b+d$, $m_2 = a+2b$, and $m_3 = -a+b+d$.  It has been 
shown~\cite{m10-2} how this pattern is obtained from $A_4$ alone 
with the help of lepton number.

Deviations from tribimaximal mixing may be obtained for $b \neq c$. 
This will allow $\nu_1$ to mix with $\nu_3$ and $\theta_{13}$ becomes 
nonzero.  However, this same mixing will move $\theta_{12}$ to a larger 
value~\cite{m04} so that $\tan^2 \theta_{12} > 0.5$ which is not favored 
by current data.  To allow $\tan^2 \theta_{12} < 0.5$, it was 
proposed~\cite{m04} that $e = -f \neq 0$ in Eq.(10).  This is maintained 
by an assumed residual symmetry of the $\xi \Phi \Phi$ soft terms of the 
Higgs potential under which $\xi_5 \leftrightarrow -\xi_6$ and 
$\Phi_2 \leftrightarrow \Phi_3$.  As a result, the neutrino mass matrix 
under $U_{TB}$ is no longer diagonal, but is given by~\cite{m04}
\begin{equation}
{\cal M}_\nu^{(1,2,3)} = \pmatrix{m_1 & 0 & m_4 \cr 0 & m_2 & m_5 \cr 
m_4 & m_5 & m_3},
\end{equation}
where $m_1 = a - (b+c)/2 + d$, $m_2 = a + b + c$, $m_3 = -a + (b+c)/2 + d$, 
$m_4 = \sqrt{3}/2 (c-b)$ and $m_5 = -i \sqrt{2} e$.  If $m_4 = 0$, 
then $\nu_2$ mixes with $\nu_3$ and it can be shown 
that the allowed range of $\theta_{23}$ from Eq.(3) implies 
$\sin^2 2 \theta_{13} \leq 0.04$ which lies on the outer edge of the 
allowed region of Eq.(5).  In the following, we consider both $m_4$ and 
$m_5$ to be nonzero and study various numerical solutions to the 
T2K data.

The atmospheric neutrino mixing is assumed to be maximal, i.e. 
$\sin^2 2 \theta_{23} = 1$, which is also the assumption of T2K 
in obtaining their new result.  The solar neutrino mixing is taken to 
be $\sin^2 2 \theta_{12} = 0.87 \pm 0.3$ ~\cite{pdg10}.  We also use 
$\Delta m^2_{32} = 2.40 \times 10^{-3}$ eV$^2$ which is the value used by T2K, 
and $\Delta m^2_{21} = 7.65 \times 10^{-5}$ eV$^2$.  For the central value 
of $\theta_{12} = 34.43^\circ$, we have $\tan^2 \theta_{12} = 0.47$ which 
is rather close to the tribimaximal prediction of 0.5.  Using this and 
assuming the central value of $\sin \theta_{13} = 0.168$ 
($\sin^2 \theta_{13} = 0.11$), the zero entry of the neutrino mass 
matrix of Eq.(12) implies the condition
\begin{equation}
0.007655 m'_1 - 0.020990 m'_2 + 0.013342 m'_3 = 0,
\end{equation}
where $m'_{1,2,3}$ are the mass eigenvalues of Eq.(12).  Hence they are 
related to the measured $\Delta m^2_{32}$ and $\Delta m^2_{21}$ by
\begin{eqnarray}
m'_2 &=& \pm \sqrt{{m'_1}^2 + \Delta m^2_{21}}, \\ 
m'_3 &=& \pm \sqrt{{m'_1}^2 + \Delta m^2_{21}/2 \pm \Delta m^2_{32}}.
\end{eqnarray}
There is only one solution to Eq.(13), i.e.
\begin{equation}
m'_1 = 0.0246~{\rm eV}, ~~~ m'_2 = -0.0261~{\rm eV}, ~~~ 
m'_3 = -0.0552~{\rm eV},
\end{equation}
which exhibits normal mass hierarchy.  From this solution, we then obtain 
$m_{1,2,3,4,5}$ and the original $A_4$ parameters $a,b,c,d,e$.
The $\nu_e$ mass observed in nuclear beta decay is given by 
$\sum_i |U_{ei}|^2 |m'_i| = 0.026$ eV. The effective mass $m_{ee}$ for 
neutrinoless double beta decay is 
\begin{equation}
m_{ee} = |a + (2/3)d| = |(2/3) m_1 + (1/3) m_2|.
\end{equation}
We plot in Figs.~1 to 3 the solutions for $|m'_{1,2,3}|$ and $m_{ee}$ 
as a function of $\sin^2 2 \theta_{13}$ in the range 0.03 to 0.135 
[corresponding to the upper bound given in Eq.(4)] for 
$\sin^2 \theta_{23} = 1$ and the values $\sin^2 2 \theta_{12} = 
0.84, 0.87, 0.90$.  Thus $m_{ee}$ is predicted to be at most 0.04 eV. 
As for the $\nu_e$ mass in nuclear beta decay, it can be read off as 
approximately given by $(2|m'_1| + |m'_2|)/3$. 
We also plot in Fig.~4 the values of $m_{1,2,3,4,5}$ for 
$\sin^2 2 \theta_{12} = 0.87$.  This shows that $m_4$ and $m_5$ , i.e. the 
parameters of $A_4$ which deviate from tribimaximal mixing, are 
indeed small.  In terms of $A_4$ symmetry, the following breaking patterns 
are in effect: in the charged-lepton sector, $A_4$ breaks to $Z_3$ 
(which may be verifed experimentally from Higgs-boson decay~\cite{cdmw11}); 
in the neutrino sector, $A_4$ breaks first to $Z_2$ (the tribimaximal limit), 
and then $Z_2$ is also broken with the pattern $b \neq c$ and $f=-e$, which may 
be maintained by a suitably chosen set of soft terms.

In conclusion, on the strength of the recent observation~\cite{t2k11} 
of a nonzero $\theta_{13}$ for neutrino mixing, the original $A_4$ 
proposal of 2004~\cite{m04} is updated.  We find that solutions are 
indeed possible with the most recent data but only in a normal 
hierarchy of neutrino masses, i.e. $|m'_1| < |m'_2| < |m'_3|$. 
We confirm that the parameters of $A_4$ which deviate from tribimaximal 
mixing, i.e. $m_4$ and $m_5$, are indeed small.  We also make 
predictions on the effective $m_{ee}$ in neutrinoless double beta decay. 

This work is supported in part by the U.~S.~Department of Energy under Grant 
No. DE-FG03-94ER40837.

\begin{figure}[htb]
\begin{center}
\includegraphics[width=1.0\textwidth]{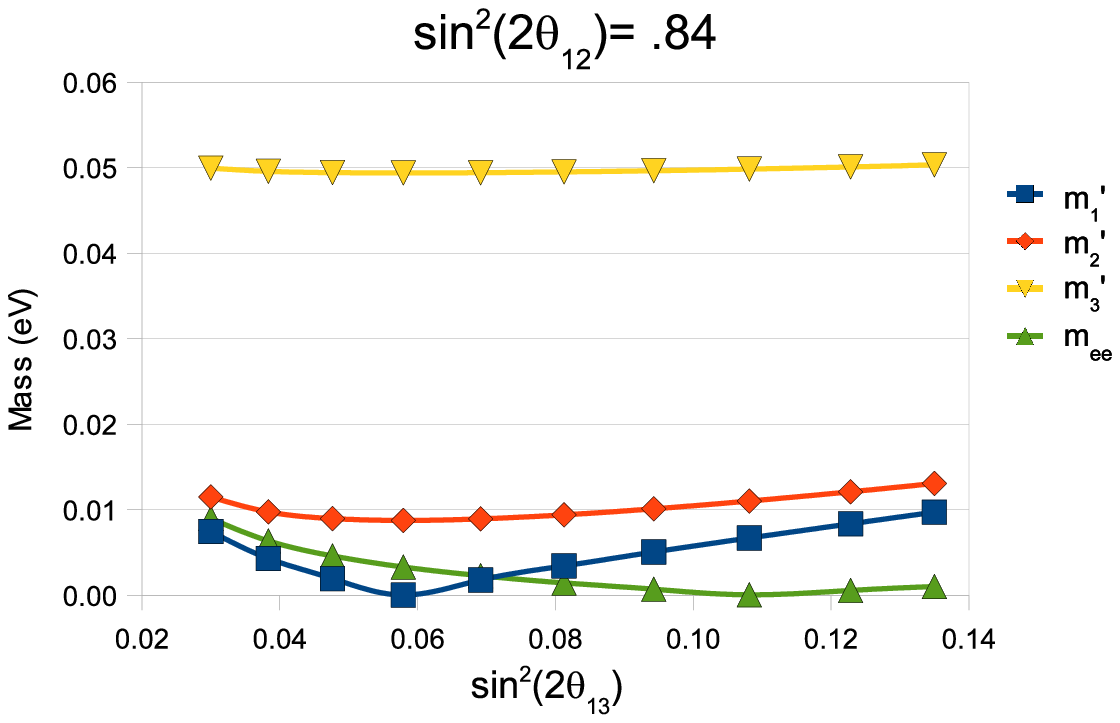}
\caption{Physical neutrino masses $|m'_{1,2,3}|$ and the effective $m_{ee}$ for 
neutrinoless double beta decay of this model in the range $0.03 \leq 
\sin^2 2 \theta_{13} \leq 0.135$ for $\sin^2 2 \theta_{23} = 1$ and 
$\sin^2 2 \theta_{12} = 0.84$.}
\end{center}
\end{figure}
\begin{figure}[htb]
\begin{center}
\includegraphics[width=1.0\textwidth]{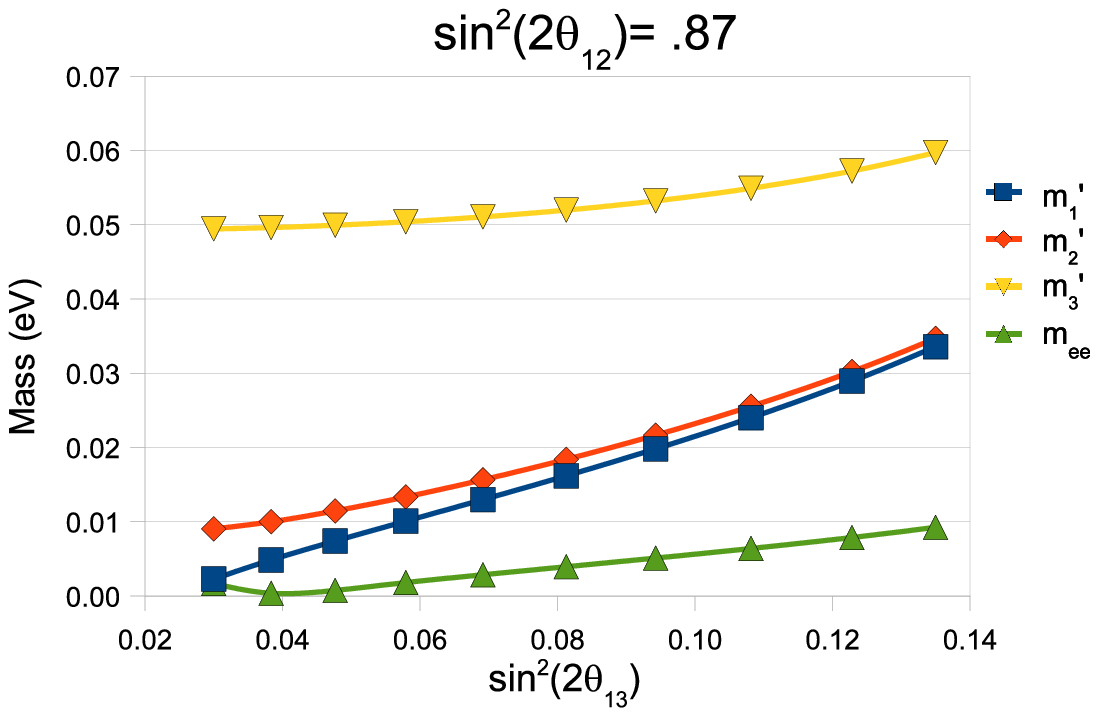}
\caption{Physical neutrino masses $|m'_{1,2,3}|$ and the effective $m_{ee}$ for 
neutrinoless double beta decay of this model in the range $0.03 \leq 
\sin^2 2 \theta_{13} \leq 0.135$ for $\sin^2 2 \theta_{23} = 1$ and 
$\sin^2 2 \theta_{12} = 0.87$.}
\end{center}
\end{figure}
\begin{figure}[htb]
\begin{center}
\includegraphics[width=1.0\textwidth]{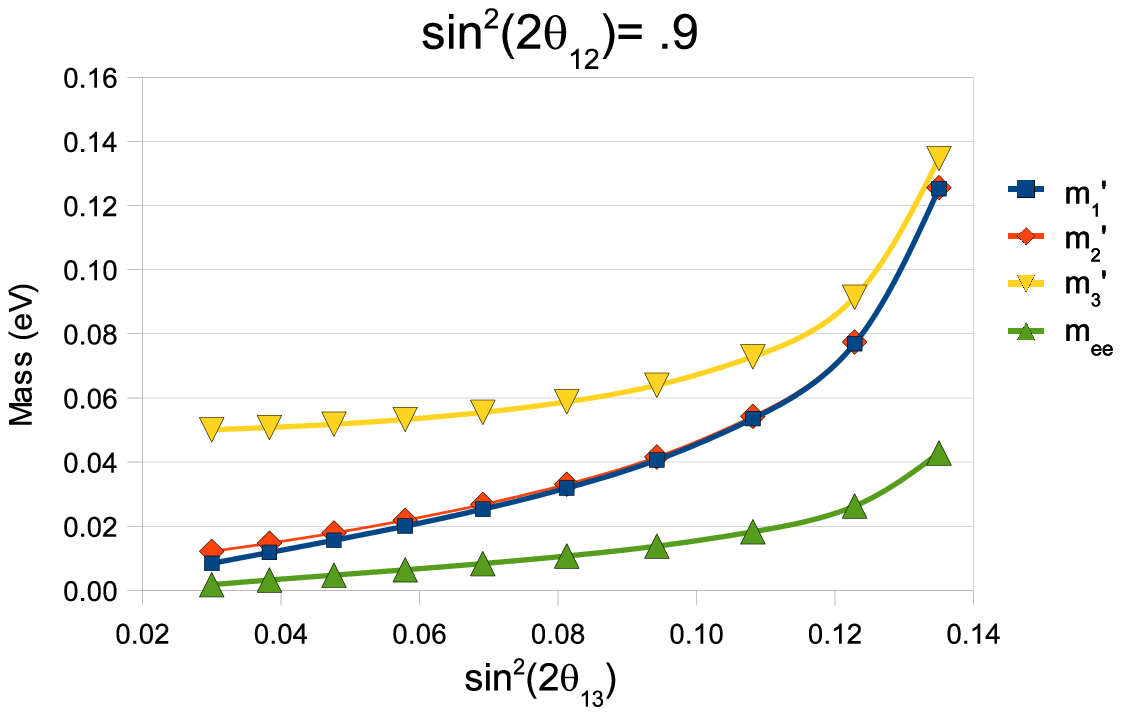}
\caption{Physical neutrino masses $|m'_{1,2,3}|$ and the effective $m_{ee}$ for 
neutrinoless double beta decay of this model in the range $0.03 \leq 
\sin^2 2 \theta_{13} \leq 0.135$ for $\sin^2 2 \theta_{23} = 1$ and 
$\sin^2 2 \theta_{12} = 0.90$.}
\end{center}
\end{figure}
\begin{figure}[htb]
\begin{center}
\includegraphics[width=1.0\textwidth]{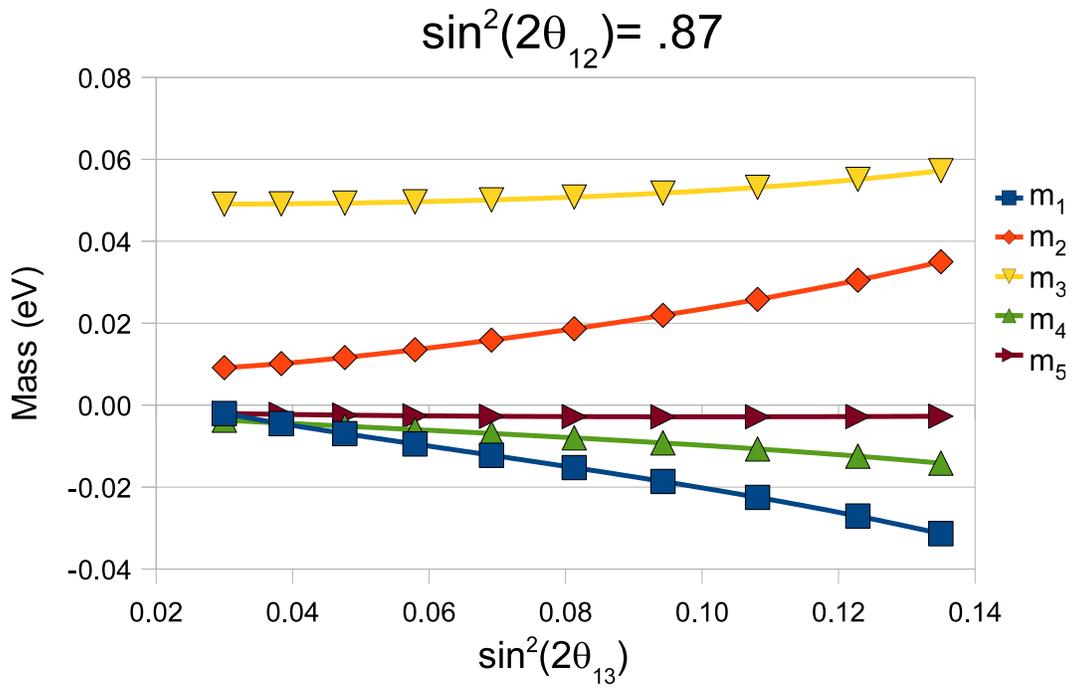}
\caption{The $A_4$ parameters $m_{1,2,3,4,5}$ of this model in the range 
$0.03 \leq \sin^2 2 \theta_{13} \leq 0.135$ for $\sin^2 2 \theta_{23} = 1$ and 
$\sin^2 2 \theta_{12} = 0.87$.}
\end{center}
\end{figure}

\end{document}